\documentclass[iop]{emulateapj}

\newcommand{\tcmbo}{\hbox{$T_\mathrm{CMB}^{\,z=0}$}}
\newcommand{\tcmb}{\hbox{$T_\mathrm{CMB}$}}
\newcommand{\tdusto}{\hbox{$T_\mathrm{dust}^{\,z=0}$}}
\newcommand{\tdust}{\hbox{$T_\mathrm{dust}$}}

\def\ts     {\thinspace} 

\def\lbg {MACS0647-JD\ts\ts}
\def\sdf1 {SDF J132415.7+273058\ts\ts}
\def\sdfa {SDF J132408.3+271543\ts\ts}
\def\lya {Ly$\alpha$\ts\ts}

\def\kms    {\ifmmode{{\rm \ts km\ts s}^{-1}}\else{\ts km\ts s$^{-1}$\ts}\fi}
\def\msol   {\ifmmode{{\rm M}_{\odot}}\else{M$_{\odot}$\ts}\fi}
\def\lsun   {\ifmmode{{\rm L}_{\odot}}\else{L$_{\odot}$\ts}\fi}
\def\cii  {\ifmmode{{\rm [C}{\rm \small II}]}\else{[C\ts {\scriptsize II}]\ts}\fi}
\def\m    {\ifmmode{\mu {\rm m}}\else{$\mu$m}\fi}
\def\hi   {\ifmmode{{\rm H}{\rm \small I}}\else{H\ts {\scriptsize I}\ts}\fi}
\def\hii  {\ifmmode{{\rm H}{\rm \small II}}\else{H\ts {\scriptsize II}\ts}\fi}
\def\hh     {\ifmmode{{\rm H}_2}\else{H$_2$\ts}\fi}
\def\nhh     {\ifmmode{N({\rm H}_2)}\else{$N$(H$_2$)\ts}\fi}
\def\microns {\ifmmode{\mu{\rm m}}\else{$\mu$m\ts}\fi}

\shorttitle{Search for [CII] emission in $z=6.5-11$  star-forming galaxies.}
\shortauthors{Gonz\'alez-L\'opez et. al.}

\begin{document}
\submitted{Accepted for publication in the Astrophysical Journal.}
\title{Search for [CII] emission in $z=6.5-11$  star-forming galaxies.}

\author{Jorge Gonz\'alez-L\'opez\altaffilmark{1,3,9}}
\author{Dominik A. Riechers\altaffilmark{2}}
\author{Roberto Decarli\altaffilmark{3}}
\author{Fabian Walter\altaffilmark{3}}
\author{Livia Vallini\altaffilmark{4}}
\author{Roberto Neri\altaffilmark{5}}
\author{Frank Bertoldi\altaffilmark{6}}
\author{Alberto D. Bolatto\altaffilmark{7}}
\author{Christopher L. Carilli\altaffilmark{8}}
\author{Pierre Cox\altaffilmark{5}}
\author{Elisabete da Cunha\altaffilmark{3}}
\author{Andrea Ferrara\altaffilmark{4}}
\author{Simona Gallerani\altaffilmark{4}}
\author{Leopoldo Infante\altaffilmark{1,9}}

\altaffiltext{1}{Instituto de Astrof\'isica, Facultad de F\'isica,  Pontificia Universidad Cat\'olica de Chile, Av. Vicu\~na Mackenna 4860, 782-0436 Macul, Santiago, Chile; jgonzal@astro.puc.cl}
\altaffiltext{2}{Astronomy Department, Cornell University, 220 Space Sciences Building, Ithaca, NY 14853, USA}
\altaffiltext{3}{Max Planck Institute f\"ur Astronomie Heidelberg, K\"onigstuhl 17, D-69117 Heidelberg, Germany}
\altaffiltext{4}{Scuola Normale Superiore, Piazza dei Cavalieri 7, I-56126 Pisa, Italy}
\altaffiltext{5}{Institut de RadioAstronomie Millim\'etrique, 300 Rue de la Piscine, Domaine Universitaire, 38406 Saint Martin d'Heres, France}
\altaffiltext{6}{Argelander Institute for Astronomy, University of Bonn, Auf dem H\"ugel 71, 53121 Bonn, Germany}
\altaffiltext{7}{Department of Astronomy, University of Maryland, College Park, MD 20742, USA}
\altaffiltext{8}{National Radio Astronomy Observatory, P. O. Box 0, Socorro, NM 87801, USA}
\altaffiltext{9}{Centro de Astro-Ingenier\'ia, Pontificia Universidad Cat\'olica de Chile, V. Mackenna 4860, Santiago, Chile}

\begin{abstract}
We present the  search for the \cii emission line in three $z>6.5$ Lyman-alpha emitters (LAEs) and
one J-Dropout galaxy  using the Combined Array for Research in Millimeter-wave Astronomy
(CARMA) and the Plateau de Bure Interferometer (PdBI). We observed three bright $z\sim6.5-7$
LAEs discovered in the SUBARU deep field (SDF) and the Multiple Imaged lensed $z\sim 11$ galaxy
candidate found behind the galaxy cluster MACSJ0647.7+7015. For the LAEs IOK-1 ($z=6.965$),
\sdf1 ($z=6.541$) and \sdfa ($z=6.554$) we find upper limits for the \cii line luminosity of
$<2.05$, $<4.52$ and $<10.56\times10^{8}\lsun$ respectively.
We find upper limits to the FIR luminosity of the galaxies using a spectral energy distribution
template of the local galaxy NGC 6946 and taking into account the effects of the Cosmic Microwave
Background on the mm observations. 
For IOK-1, \sdf1 and \sdfa we find upper limits for the FIR luminosity of $<2.33$, $3.79$ and
$7.72\times10^{11}\lsun$ respectively. For the lensed galaxy \lbg, one of the highest redshift
galaxy candidate to date with $z_{\rm ph}=10.7^{+0.6}_{-0.4}$ we put an upper limit in the \cii
emission of $<1.36\times10^{8}\times(\mu/15)^{-1}\lsun$ and an upper limit in the FIR luminosity
of  $<6.1\times10^{10}\times(\mu/15)^{-1}\lsun$ (where $\mu$ is the magnification factor). 
We explore the different conditions relevant for the search for \cii emission in high redshift galaxies
as well as the difficulties for future observations with ALMA and CCAT.
\end{abstract}



\section{Introduction}

Lyman-alpha Emitters (LAE) are galaxies selected through strong \lya emission and are among the
most studied galaxy populations at high redshift. The use of narrow band filters over a wide area on
the sky has proven to be a very effective method to find galaxies up to $z\sim7$
\citep{iye2006,fontana2010,vanzella2011,rhoads2012,shibuya2012,schenker2012,ono2012}. The
possibility of finding LAEs from $z\sim1$  to $z\sim7$  shows that this type of galaxies can be
used to understand galaxy evolution over cosmic time. 
It has been observed that the LAE fraction in UV selected galaxies increases with redshift up to 
$z\sim6$ \citep{stark2010}, which is expected due to the decreasing dust content at higher
redshifts. Beyond $z\sim6$ it is expected that the LAE fraction decreases as the amount of neutral
Hydrogen (\hi) increases, due to the incomplete reionization of the Intergalactic Medium (IGM)
\citep{ota2008,stark2010,pentericci2011,ono2012,schenker2012}. This is consistent with the
comparatively low success rate of detection of \lya emission at $z\gtrsim 7$.

If Lyman-alpha photons from redshifts $z\geq7$ are absorbed by \hi\ts in the IGM
\citep{dayal2012}, it will be difficult to spectroscopically confirm the candidates at high redshift, 
such as the candidate $z\sim12$ galaxy UDFj-39546284 discovered in the Hubble Space Telescope
(HST) Ultra Deep Field (UDF) \citep{bouwens2011a,ellis2013,brammer2013,capak2013}, and the
candidates found behind galaxy clusters at $z\sim9.6$ MACS1149-JD  and $\sim10.7$ \lbg
\citep{zheng2012,coe2013}.

Among the usual Interstellar medium (ISM) tracers at optical/UV wavelengths, the only line that has
been observed at $z>4$ in galaxies is Ly-alpha. The emission of Ly-alpha is complicated by its high
optical depth in the emission region and its escape through resonant scattering, by dust absorption,
and by the contribution from outflows. 
Therefore direct constrains on the gas properties from the Ly-alpha line strength and shape are
difficult to derive. This motivates the exploration of alternative means to study the highest redshift
galaxies. Promising candidates include far-infrared fine structure emission lines, e.g., \cii 
($^2$P$_{3/2}$ $\rightarrow$ $^2$P$_{1/2}$) at $157.74$ $\mu$m, which is not affected by the
increasingly neutral IGM at $z>7$ and can account for up to $1\%$ of the total infrared luminosity
in some galaxies, especially in those with low luminosity and metallicity 
\citep{crawford1985,stacey1991,israel1996,madden1997}.

The \cii line  traces photo-dissociation (a.k.a. photon-dominated) regions (PDRs), as well as diffuse
\hi\ts and \hii \ts  regions. In PDRs, the far-UV radiation produced by OB stars heats the surface
layers of molecular clouds, which cool preferentially  through \cii emission. It has been observed
that most of the \cii emission in IR-bright galaxies is coming from PDRs, and that the PDR gas mass
fraction can be up to $50\%$ in starbursts like M82 \citep{crawford1985}.

Modeling of FIR emission lines observed in starburst galaxies showed that at least 70\% of the \cii
emission is produced in PDRs \citep{carral1994,lord1996,colbert1999}. In the low-metallicity
system Haro 11, on the other hand, at least 50\% of the \cii emission arises from a more diffuse,
extended ionized medium \citep{cormier2012}. The different conditions in which the \cii emission
is produced, and the direct or indirect relation of these conditions with the star formation process, 
suggest that  \cii emission should be a good tracer of the global galactic star formation activity
\citep{delooze2011}, at least for galaxies with low $T_{\rm dust}$ or low 
$\Sigma_{\rm IR}=L_{\rm IR}/\pi r^{2}_{\rm mid-IR}$ \citep{diaz-santos2013}. \cii is found to be
the strongest emission line, stronger than CO, and thus is the most promising tracer of the dense,
star forming regions in distant galaxies where \cii can be detected with ground-based telescopes
due to the redshift into observable atmospheric windows.

In the past years, the \cii 158 $\mu$m emission line has been established as a promising 
observable in high-redshift galaxies \citep{maiolino2005,
iono2006,maiolino2009,walter2009,hailey-dunsheath2010,stacey2010,ivison2010,wagg2010,
cox2011,debreuck2011,valtchanov2011,gallerani2012,venemans2012,wagg2012,walter2012a,carilli2013,wang2013,willott2013,riechers2013}. 
Most of the high-z detections were for infrared-luminous starbursts, many of which also show signatures of AGN. See the review by \cite{carilli_walter2013} for more details.

With star formation rates of a few tens $\msol \rm{yr}^{-1}$, based on the \lya and UV continuum emission, LAEs are classified as ``normal''
star forming galaxies. Different studies claim that LAEs are young, dust free, 
starbursting galaxies, supported by UV observations \citep{gawiser2006,finkelstein2007,lai2008}. Recent MIR detection
of LAEs at $z\sim2.5$ and $z<0.3$ show that a significant fraction of the star formation
in these galaxies is strongly obscured by dust \citep{oteo2012a,oteo2012b}. Thus, LAEs are promising targets for the 
detection of \cii at high redshift.

Previous attempts to detect \cii in a small sample of LAEs at $z\sim6.6$ were unsuccessful
 \citep{walter2012,kanekar2013,ouchi2013}.

Here we present the result of a search for \cii in three LAEs at $z>6.5$ and in 
a lensed galaxy at $z\sim11$. In Sect. 2 we describe the target selection and 
observations. The data is shown in Sect. 3 together with some implications 
and analysis in Sect. 4. A summary of the paper is presented in Sect. 5. 
Throughout this paper we use a $\Lambda$-Cold Dark Matter cosmology with 
H$_{0}$ = 70 km s$^{-1}$ Mpc$^{-1}$, $\Omega_{\Lambda}=0.7$ and 
$\Omega_{\rm m}=0.3$.

\section{Observations}
\subsection{Source Selection}
The three Lyman-$\alpha$ emitters targeted in this study were discovered in the Subaru Deep Field (SDF). 
Two of the LAEs observed belong to the sample of LAEs at
$z\sim6.6$ discovered by \cite{taniguchi2005}. 
The targets are the brightest LAEs (sources 3 and 4 in their catalog) and 
have a narrow and bright Lyman-$\alpha$ emission line. 
The third LAE (IOK-1) was discovered at $z\sim 7$ by \cite{iye2006}. It is one of the
brightest and most distant LAEs known to date.

The fourth target, MACS0647-JD, is a lensed Lyman-break galaxy (LBG)
discovered behind the galaxy cluster MACSJ0647.7+7015 at $z=0.591$ 
\citep{coe2013}. The galaxy was discovered as a J-Dropout galaxy lensed 
into 3 magnified images  as part of The Cluster Lensing And Supernova 
survey with Hubble (CLASH) \citep{postman2012}. The three images of 
the galaxy MACS0647-JD1, MACS0647-JD2 and MACS0647-JD3, have a 
magnification of $\sim$8,$\sim$7 and $\sim$2 respectively. The photometric 
redshift of the galaxy is $10.7^{+0.6}_{-0.4}$ (95\% confidence limits).  
This is one of the highest redshift galaxy candidates known to date.

\subsection{CARMA Observations}
Observations of the three $z\sim6.5-7$ LAEs were carried out using the Combined Array for Research in 
Millimeter-wave Astronomy (CARMA) between 2008 July and 2010 July. 
The array configurations used were the most compact, D and E, to minimize phase decoherence and maximize point source sensitivity. 
The \cii line has a rest frequency of 1900.54 GHz (157.74 $\mu$m). For 
the redshifts of the targets, the line is shifted to the 1 mm band. The receivers 
were tuned to a  frequency $\sim$150 
\kms bluer than the expected frequency from the redshift 
determined by the peak of the Lyman-$\alpha$ line. This is for taking into account 
the possible absorption by the IGM in the Lyman-$\alpha$ line. 
The setups provide an instantaneous bandwidth of $\sim1.5$ GHz ($\sim 1800 \kms$) with a 
spectral resolution of 31.25 MHz ($\sim37-39$ \kms).

The observations were processed using MIRIAD \citep{sault1995}. The absolute 
flux calibrators used  are 3C84, MWC349, 3C273 and Mars, the latter being 
the most used. As passband calibrators the QSOs 3C273, 3C345 
and 0854+201 were used. As gain calibrator the QSO 1310+323 was used.
The time on source for IOK-1 was 58.5 hours, for \sdf1 was 15.9 hours and for \sdfa 4.6 hours.
The final cubes were made using natural weighting to maximize point 
source sensitivity. The observations resulted in the following beamsizes: IOK-1: 
$1.86" \times 1.33"$, PA$=-0.34^{\circ}$, SDF J132415.7+273058: 
$1.92" \times 1.56"$, PA$=83.45^{\circ}$, SDF J132408.3+271543: 
$2.54" \times 2.01"$, PA$=88.02^{\circ}$ (all targets: D and E configurations). 
For D configuration the minimum baseline is 11 meters and the maximum is 150 meters.
For E configuration the minimum baseline is 8 meters and the maximum is 66 meters. 
Table \ref{tab:table1} summarizes the sensitivities reached for the observations of the LAEs.

\subsection{PdBI Observations}
All \lbg observations were carried out in 2012 November as part of a DDT (Director's 
Discretionary Time) program with the Plateau de Bure Interferometer (PdBI). The 
target was observed with 4 WideX frequency setups (3.6 GHz bandwidth each), covering 80\% 
of the photometric redshift range ($z=10.1-11.1$). Two of the three lensed images (JD1 and JD2) are within $18"$ of each other and 
they were covered in a common 2 mm pointing.
The absolute flux calibrators used  are MWC349, 2200+420, 3C279 and 0716+714. As gain calibrator the QSO 0716+714 was used.
The total on source time for all tunings was 7.4 hours (6-antennas equivalent). The observations were processed  using GILDAS.
The beamsize of the observations is the following: \lbg: $2.10" \times 1.76"$, PA$=102.0^{\circ}$ (C configuration).
For C configuration the minimum baseline is 22 meters and the maximum is 184 meters. 
Table \ref{tab:table2} summarizes the sensitivity reached for the observations of \lbg.

\section{Results}
\begin{deluxetable*}{ccccccccccc}
\tabletypesize{\scriptsize}
\tablecaption{Summary of Observations and Results for the LAEs\label{tab:table1}}
\tablehead{
\colhead{source} & \colhead{RA} & \colhead{DEC}  & \colhead{z\tablenotemark{a}} & \colhead{$\nu_{\rm obs}$\tablenotemark{b}} & \colhead{$\sigma_{\rm cont}$\tablenotemark{c}} & \colhead{$\sigma_{\rm line}$\tablenotemark{d}}  & \colhead{L$_{\rm [CII]}$\tablenotemark{e}} & \colhead{L$_{\rm IR, CMB}^{\rm N6946}$\tablenotemark{f}} & \colhead{SFR$_{\rm dust, CMB}$\tablenotemark{g}} & \colhead{SFR$_{\rm UV}$\tablenotemark{h}}\\
\colhead{}& \colhead{J2000.0}  & \colhead{J2000.0}&  \colhead{}   & \colhead{GHz} & \colhead{mJy\,b$^{-1}$}   &  \colhead{mJy\,b$^{-1}$} &  \colhead{10$^8$\,L$_{\odot}$}  &  \colhead{10$^{11}$\,L$_{\odot}$}& \colhead{M$_{\odot}$\,yr$^{-1}$} & \colhead{M$_{\odot}$\,yr$^{-1}$ }
}
\startdata
IOK--1& 13:23:59.80 & +27:24:56.0  & 6.965     & 238.881        & 0.19  & 1.17  & $<$2.05 & $<$6.34 &  $<$109.1 & $\sim$24  \\
SDF J132415.7 & 13:24:15.70 & +27:30:58.0  & 6.541     & 252.154        & 0.37  & 2.82  & $<$4.52 & $<$10.3 &  $<$177.2 & $\sim$34  \\
SDF J132408.3 & 13:24:08.30 & +27:15:43.0  & 6.554     & 251.594        & 0.75  & 5.67  & $<$10.56 & $<$21.0 &  $<$360.9 & $\sim$15 
\enddata
\tablecomments{All luminosities upper limits are 3\,$\sigma$.}
\tablenotetext{a}{References: IOK--1:\cite{iye2006,ono2012} --\sdf1 and \sdfa: \cite{taniguchi2005}}
\tablenotetext{b}{Observing Frequencies; tuned $\sim$125\,MHz blueward of the Ly--$\alpha$ redshifts for all targets.}
\tablenotetext{c}{1$\sigma$ continuum sensitivity at 158$\mu$m rest wavelength.}
\tablenotetext{d}{1$\sigma$ \cii\ line sensitivity over a channel width of 50 km\,s$^{-1}$.}
\tablenotetext{e}{3$\sigma$ \cii\ luminosity limit over a channel width of 50 km\,s$^{-1}$ assuming $L_{\rm line} = 1.04 \times 10^{-3} \, I_{\rm line} \, \nu_{\rm rest} (1+z)^{-1}\, D_{\rm L}^{2}$, where the line luminosity, $L_{\rm line}$, is measured in $L_\odot$; the velocity integrated flux, $I_{\rm line}$=$S_{\rm line}\, \Delta v$, in Jy\,\kms; the rest frequency, $\nu_{\rm rest} = \nu_{\rm obs} (1+z)$, in GHz; and the luminosity distance, $D_{\rm L}$, in Mpc. (e.g. \cite{solomon1992}}
\tablenotetext{f}{3$\sigma$ limit based on the SED of NGC 6946 and including the effect of the CMB.}
\tablenotetext{g}{3$\sigma$ limit based on L$_{\rm IR}^{\rm N6946}$ including the effect of the CMB.}
\tablenotetext{h}{UV--based SFR from \cite{jiang2013}}
\end{deluxetable*}
\begin{center}
\begin{deluxetable}{lc}
\tabletypesize{\scriptsize}
\tablecaption{Summary of Observations and Results for MACS0647JD\label{tab:table2}}
\tablewidth{0pt}
\tablehead{
\colhead{Parameter} & \colhead{MACS0647-JD1, JD2} 
}
\startdata
Coordinates (J2000) JD1          & 06:47:55.731,+70:14:35.76	\\
Coordinates (J2000) JD2          & 06:47:53.112,+70:14:22.94	\\
$\mu$ (JD1+JD2)						& $\sim15$									\\
Redshift  									& $10.7^{+0.6}_{-0.4}$				\\
UV SFR										& $\sim 1$ [M$_{\odot}$yr$^{-1}$]\\
$\nu$									    & 156.7-171.1 [GHz]					\\
$\sigma_{\rm cont}$ \tablenotemark{a}	    			& 0.17 mJy\,b$^{-1}$									\\
$\sigma_{\rm line}$ (Setup A)\tablenotemark{b}	& 3.31 mJy\,b$^{-1}$								\\
$\sigma_{\rm line}$ (Setup B)\tablenotemark{b}		& 4.12 mJy\,b$^{-1}$										\\
$\sigma_{\rm line}$ (Setup C)\tablenotemark{b}		& 3.19 mJy\,b$^{-1}$								\\
$\sigma_{\rm line}$ (Setup D)\tablenotemark{b}		& 6.42 mJy\,b$^{-1}$									\\
$L_{\cii}$ (Setup C)\tablenotemark{c}					& $<6.78\times10^{7}\times(\mu/15)^{-1}[\lsun]$		\\
$L_{\cii}$ (Setup D)\tablenotemark{c}					& $<1.36\times10^{8}\times(\mu/15)^{-1}[\lsun]$		\\
L$_{\rm IR}^{\rm N6946,}$\tablenotemark{d} (Corrected CMB)					& $<1.65\times10^{11}\times(\mu/15)^{-1}[\lsun]$	\\
SFR ($L_{\rm IR}$) (Corrected CMB)\tablenotemark{e}					& $<28\times(\mu/15)^{-1}$ [M$_{\odot}$yr$^{-1}$]	\\
SFR ($L_{\cii}$) (Setup C)\tablenotemark{f}					& $<5\times(\mu/15)^{-1}$ [M$_{\odot}$yr$^{-1}$]	\\
SFR ($L_{\cii}$) (Setup D)\tablenotemark{f}					& $<9\times(\mu/15)^{-1}$ [M$_{\odot}$yr$^{-1}$]	
\enddata
\tablecomments{All luminosities upper limits are 3\,$\sigma$.\\References: Coordinates, 
magnification, redshift and UV-SFR from Coe et al. 2013.\\All the luminosities and SFR are corrected by  magnification.}
\tablenotetext{a}{1$\sigma$ continuum sensitivity at 158$\mu$m rest wavelength.}
\tablenotetext{b}{1$\sigma$ \cii\ line sensitivity over a channel width of 50\,km\,s$^{-1}$.}
\tablenotetext{c}{3$\sigma$ \cii\ luminosity limit over a channel width of 50\,km\,s$^{-1}$ as in Tab. \ref{tab:table1}. The two results correspond to the most sensitive and the least sensitive setups. }
\tablenotetext{d}{3$\sigma$ limit based on the SED of NGC 6946 and including the effect of the CMB.}
\tablenotetext{e}{3$\sigma$ limit based on L$_{\rm IR}^{\rm N6946}$ including the effect of the CMB.}
\tablenotetext{f}{Based on the De Looze et al., 2011 $L_{\cii}-SFR$ relation.  The two results correspond to the most sensitive and the lest sensitive setups.}
\end{deluxetable}
\end{center}
\begin{figure}
\epsscale{1.2}
\plotone{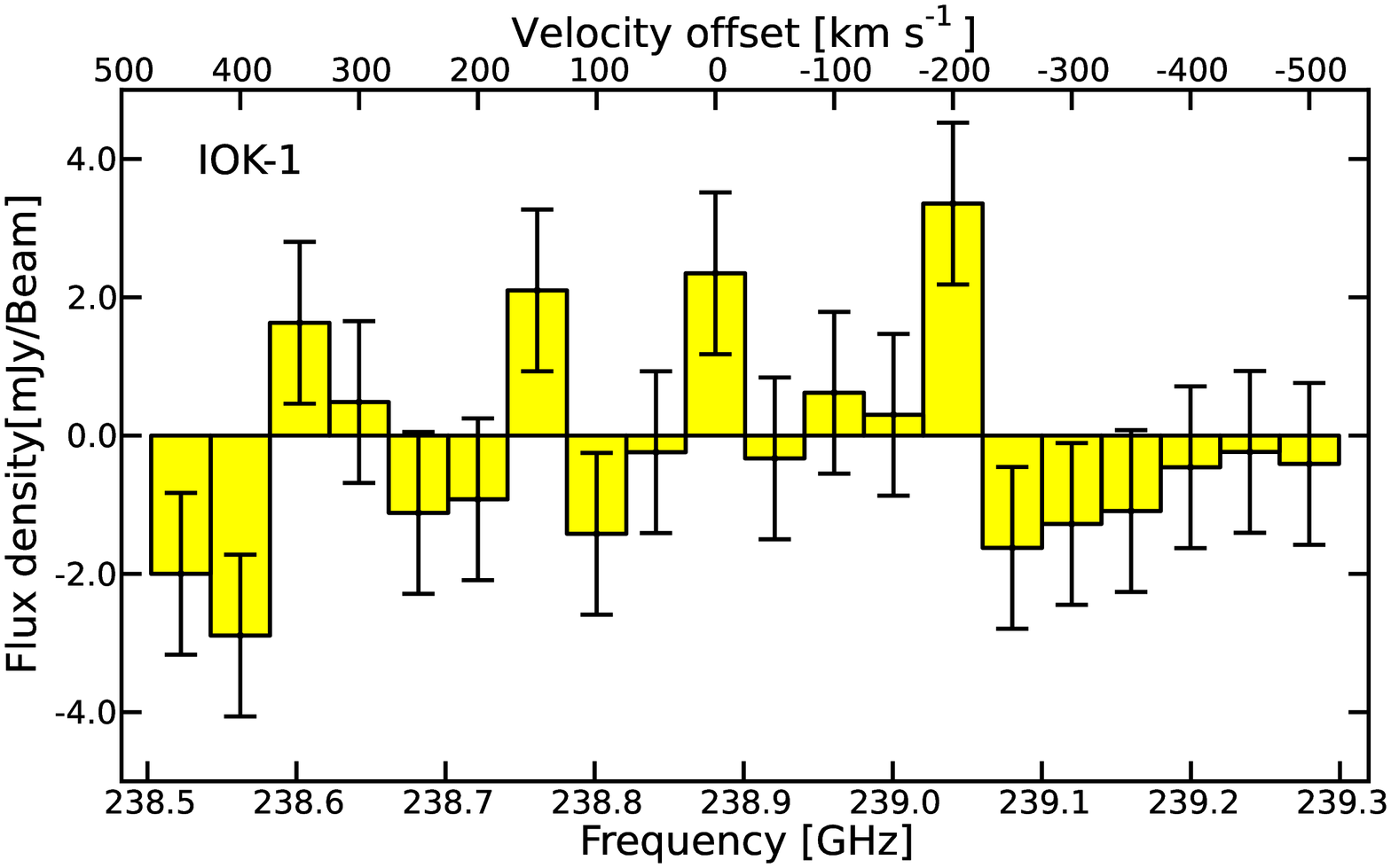}
\plotone{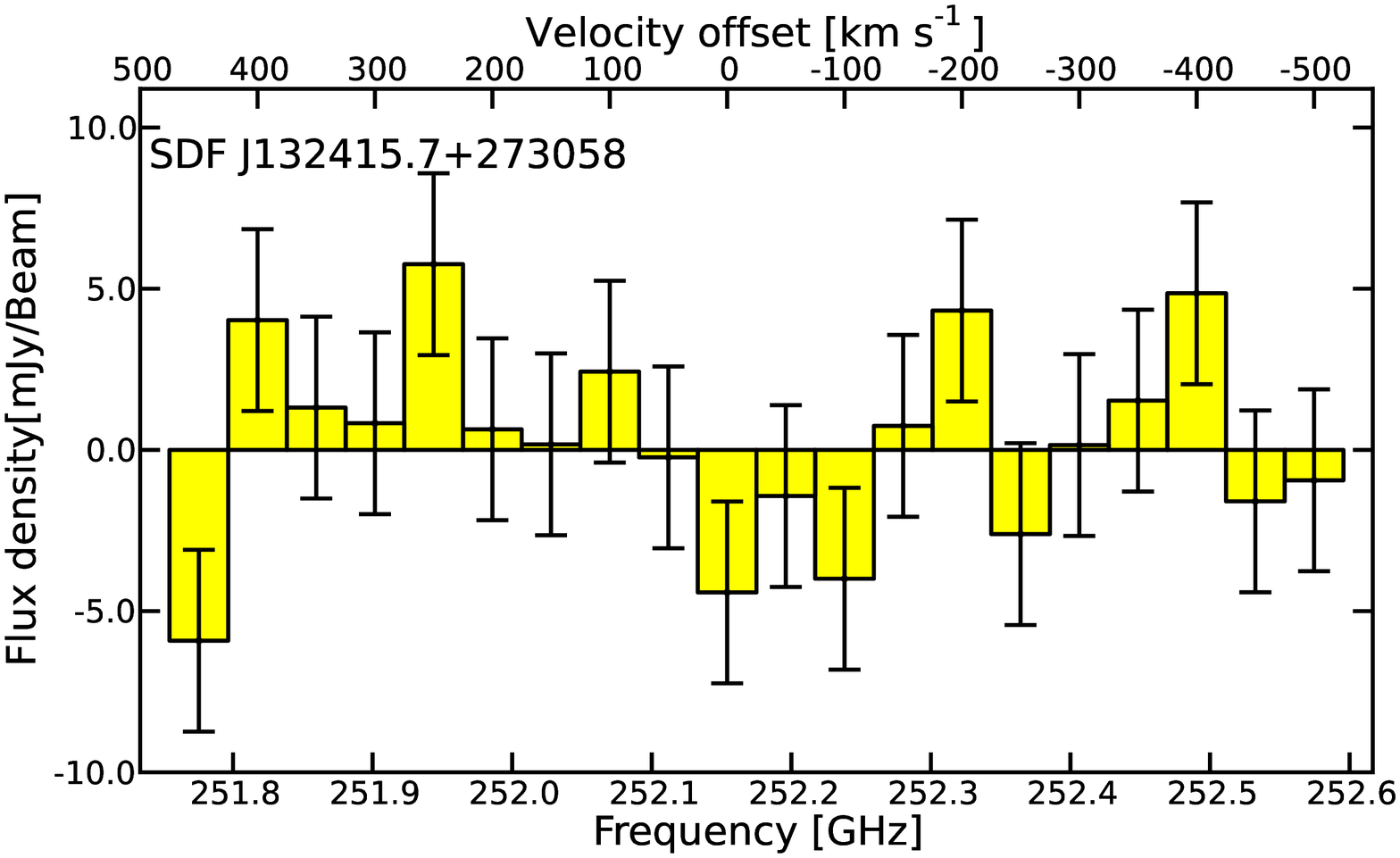}
\plotone{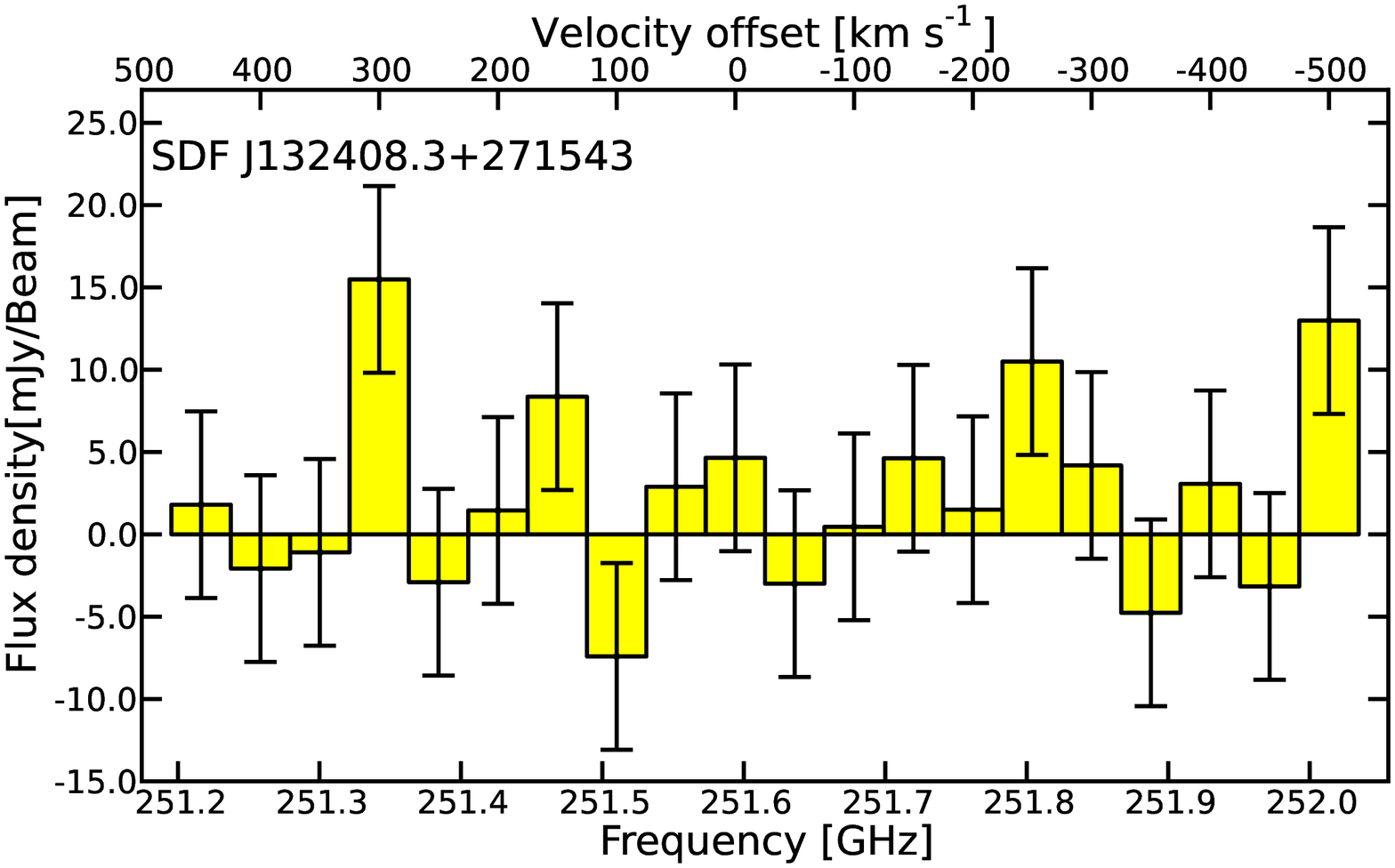}
\caption{Spectra of the LAEs with a  velocity resolution of 50 \kms. The relative velocities are with respect to the frequency expected
for the \cii line including absorption by the IGM (150 \kms to the blue of $z_{\rm Ly\alpha}$). The redshifts of the target are z=6.965 for IOK-1, z=6.541 for \sdf1 and z=6.554 for \sdfa. \label{fig:spectraLAEs}}
\end{figure}
\begin{figure*}
\epsscale{1.2}
\plotone{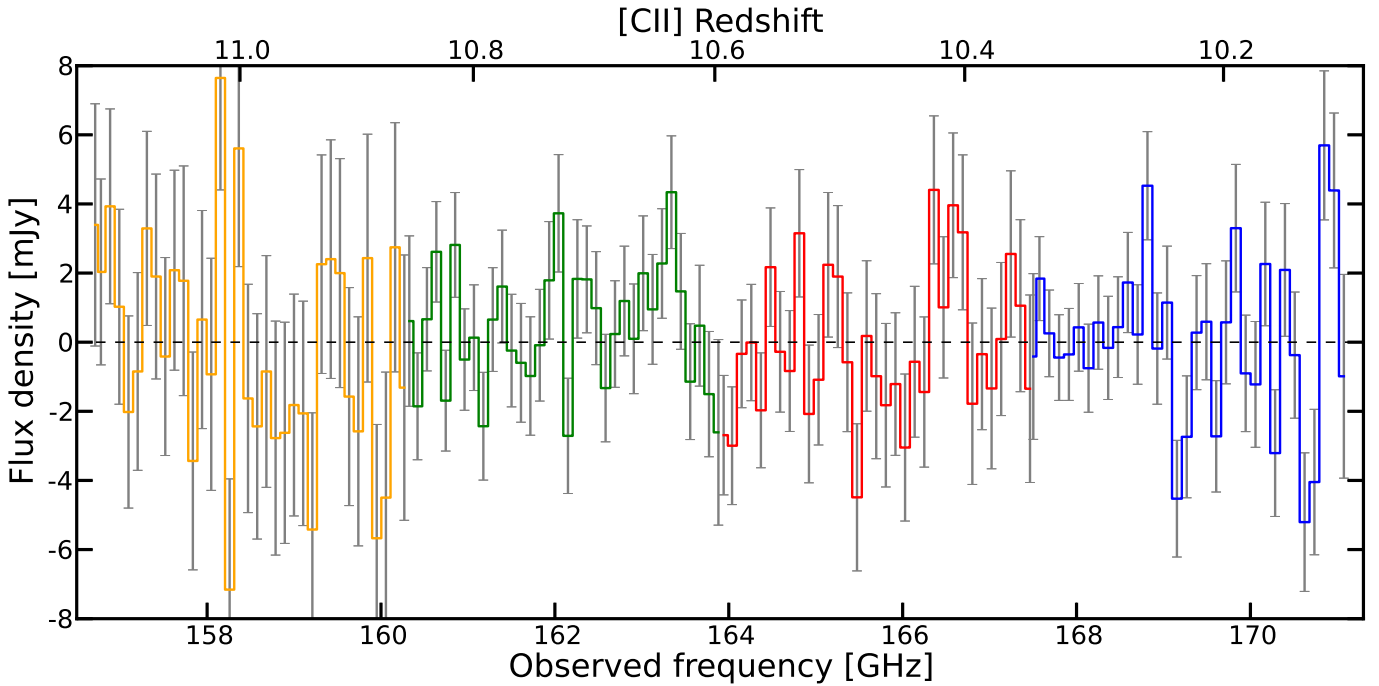}
\caption{Spectrum of \lbg. The spectrum shows the added fluxes measured on the positions of the two lensed images JD1 and JD2 (combined magnification $\mu\sim15$). The spectra of the two images were corrected by the primary beam pattern before combination.
The 4 setups are plotted in different colors, blue, red, green and orange the colors for the setups A, B, C and D respectively.
The error bars correspond to the quadrature of the errors of the individual measurement of the fluxes for JD1 and JD2 in each frequency channel. 
For display purposes, the spectrum is sampled at a channel resolution of 200 \kms, but the search of the \cii line as well as the analysis was made with the
spectrum sampled to 50 \kms. 
   \label{fig:specLBG}}
\end{figure*}
\begin{figure*}
\epsscale{1.2}
\plotone{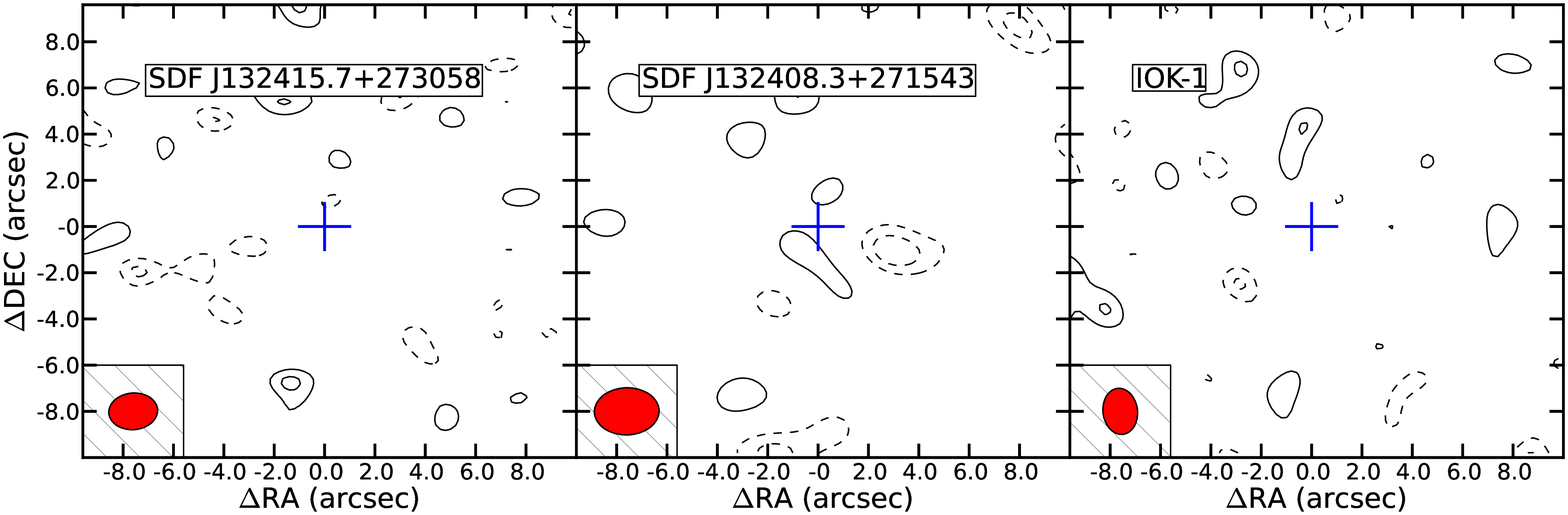}
\caption{Rest-frame 158 $\mu m$ continuum maps of the LAEs. Each contour level represents $1\sigma$ steps ($\pm1\sigma$ levels are not shown). Solid contours are positive signal 
and dashed contour are negative signals. The $1\sigma$ levels are 0.75 $\rm{mJy}\ts\rm{beam}^{-1}$ for \sdfa, 0.37 $\rm{mJy}\ts\rm{beam}^{-1}$ for \sdf1 and 0.19 $\rm{mJy}\ts\rm{beam}^{-1}$ for IOK-1.
The blue crosses represent the position of each LAE as given in Tab. \ref{tab:table1}. \label{fig:contCarma}}
\end{figure*}
\begin{figure}
\epsscale{1.2}
\plotone{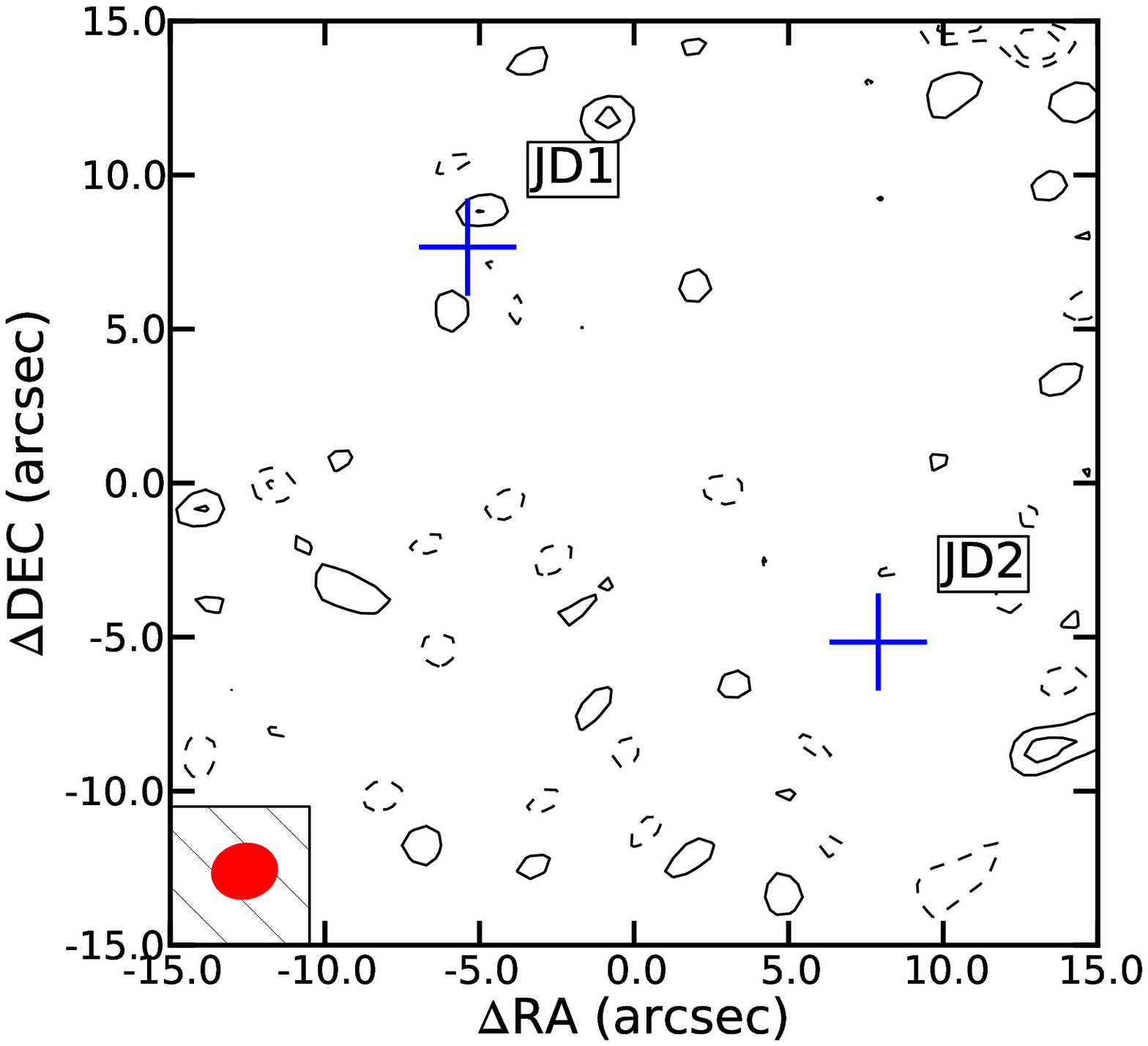}
\caption{Continuum map of the field of \lbg.  Each contour level represents $1\sigma$ steps ($\pm1\sigma$ levels are not shown). Solid contours are positive signal 
and dashed contour are negative signals. The $1\sigma$ level is 93 $\mu\rm{mJy}\ts\rm{beam}^{-1}$. The blue plus signs represent the positions of the 
two lensed images MACS0647-JD1 and MACS0647-JD2 as given in Tab. \ref{tab:table2}.
\label{fig:contLBG}}
\end{figure}
\subsection{Line Emission}

The spectra of the three $z\sim6.5-7$ LAEs are presented in Fig. \ref{fig:spectraLAEs} and the spectrum of \lbg is shown in Fig.\ref{fig:specLBG}.
No significant emission is detected at the redshifted line frequencies or close to them. 
The observations were sampled to a channel resolution of 50 \kms similar to
the expected FWHM of the \cii emission line  (see Sect. \ref{sec:widthline}). 
We use our non-detections to put constrains on the luminosities of the \cii lines 
for all targets. The results for the LAEs can be seen in the Table \ref{tab:table1}
and for the \lbg in Table \ref{tab:table2}. The upper limits were estimated assuming that the sources were unresolved. For \lbg the spectra of the two images were
corrected by the primary beam pattern before combination.
The \cii luminosities were estimated assuming that the velocity integrated flux 
of the line is $I_{\rm line}$=$S_{\rm line}\, \Delta v$, with $S_{\rm line}$ being 3 times the r.m.s. 
of the 50 \kms channel and $\Delta v=50\kms$ the range in velocity (details on Tab. \ref{tab:table1} notes). 
Using  $3\sigma$ over a 50 \kms channel to estimate the upper limit in the luminosities can result in a underestimation. 
We point out that for a  more conservative estimation the luminosities should be multiplied by a factor 2. (i.e. $3\sigma$ over
200 \kms channel). Assuming a channel width of 200 \kms, our IOK-1 \cii limit is $\sim10\%$ deeper than the 
previous PdBI limit \citep{walter2012}.

\subsection{Continuum Emission}
No continuum emission is detected in our observations of the three LAEs and 
the $z\sim 11$ LBG. The sensitivity reached for the continuum observations is 
given in the Tab. \ref{tab:table1} for the LAEs and a continuum map for the three LAEs 
is shown in Fig. \ref{fig:contCarma}.
The results for the \lbg are given in Tab. \ref{tab:table2} and the continuum 
map is shown in Fig. \ref{fig:contLBG}. 
In Sect. \ref{sec:cmbeffects} we discuss how the CMB affects our continuum observations
and in Sect. \ref{sec:sed} we use our continuum measurements to constrain the nature of our targets.

\begin{figure}
\epsscale{1.2}
\plotone{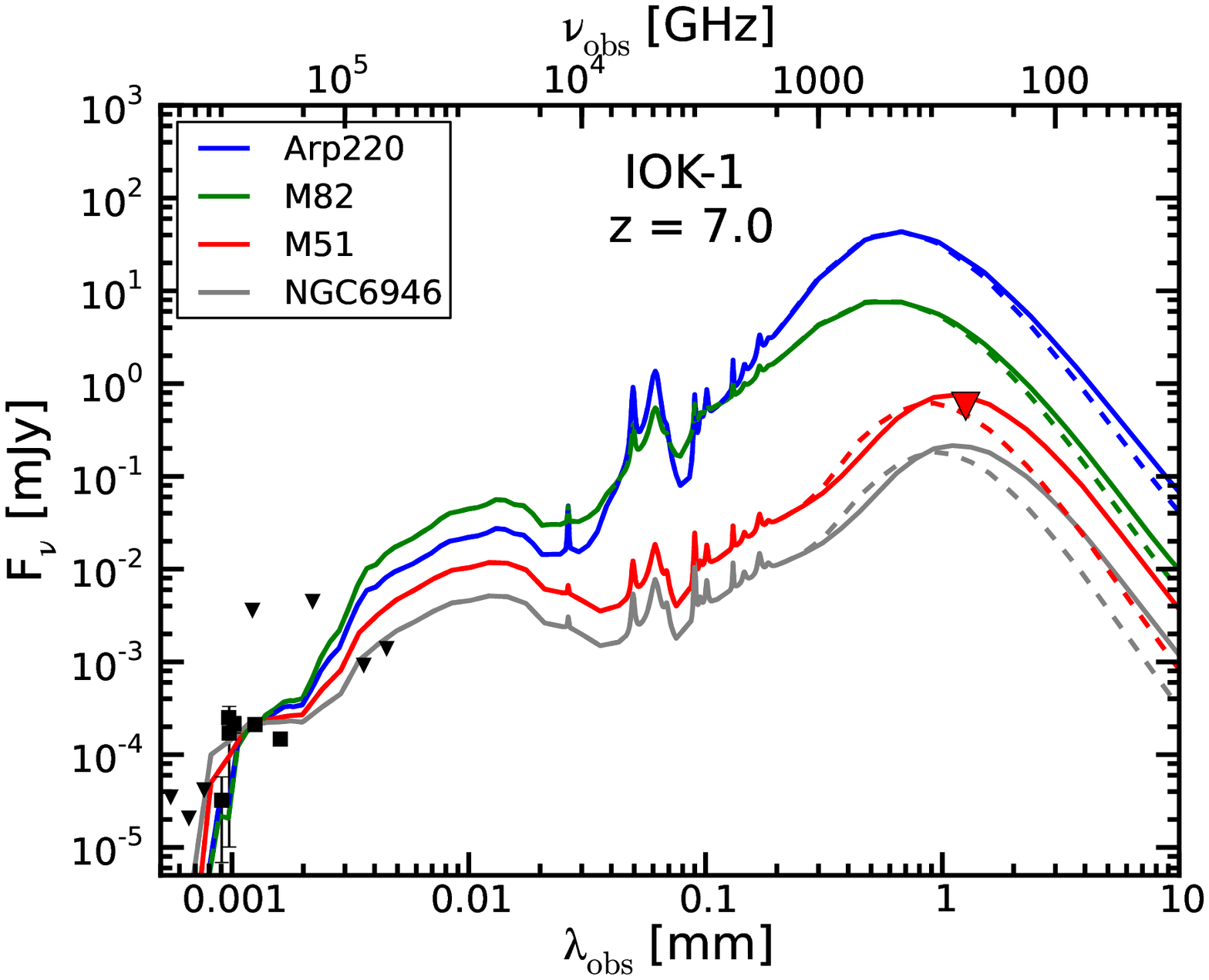}
\plotone{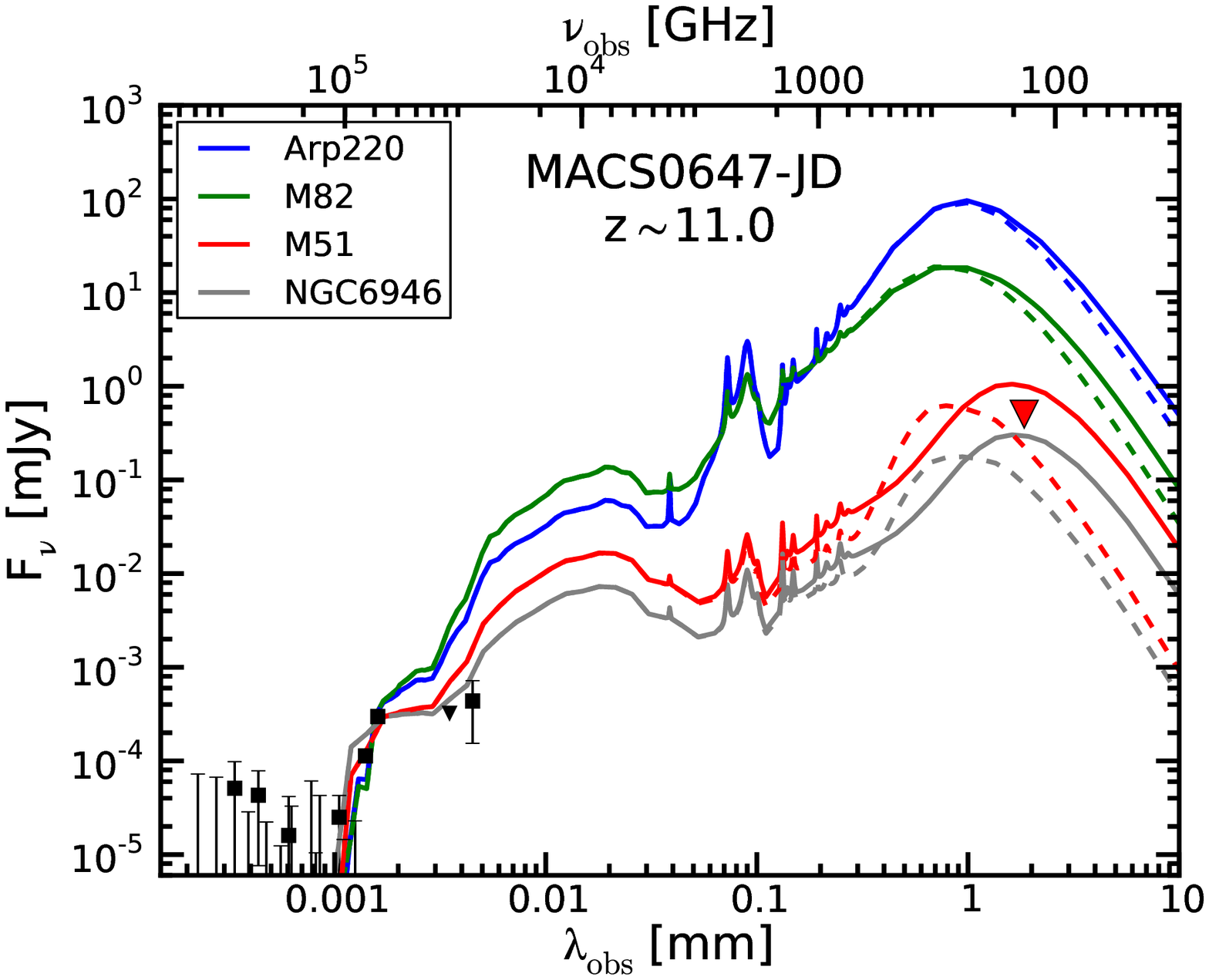}
\caption{{\it Top:} Spectral energy distribution of IOK-1. The photometric points correspond to those
measured by \cite{iye2006,ota2010,cai2011,ono2012}. The red triangle corresponds to the $3\sigma$ upper limit
given by the CARMA observations. The colored lines correspond to the SEDs of local galaxies shifted to the redshift
of IOK-1 and scaled to the observations in the UV band. The dashed lines correspond to the observed SED of the local galaxies after the effects
of the CMB on the observations are taken into account. 
{\it Bottom:} Spectral energy distribution of MACS0647-JD.  The photometric points correspond to those presented by \cite{coe2013}. 
The SED of the galaxies follow the same prescription as in the upper panel. The red triangle corresponds to the $3\sigma$ upper limit calculated as the quadrature of the errors of the individual fluxes of JD1 and JD2, in the same way as the errors presented by \cite{coe2013}. The $1\sigma$ photometric uncertainty of the observations is 0.093 mJy, and the error of the added fluxes is 0.13 mJy.
\label{fig:SEDs}}
\end{figure}

\section{Discussion}

\subsection{Width of the [CII] emission line.}
\label{sec:widthline}
Previous studies have presented the non-detection of \cii \citep{walter2012,
ouchi2013} with a channel resolution of 200 \kms, a choice motivated by the width 
of the \lya emission line. We argue that recent observations 
and models of \cii in LAEs suggest that the \cii line could be 
narrower that the previously assumed value.

\subsubsection{\cii detection on a LAE z=4.7.}

In support of a the narrow emission line is the detection of \cii 
in a LAE at $z=4.7$ (\lya-1) with ALMA \citep{carilli2013}. The FWHM of the emission 
line is 56 \kms, which is one order of magnitude narrower than the width 
of the \lya emission line of $\sim 1100$ \kms of the same source \citep{petitjean1996,ohyama2004}. 
Despite of the LAE being at a separation of 2.3$''$ ($\sim 15$ kpc) to the quasar 
BRI 1202-0725, there is no evidence for a significant influence of the quasar on the properties of the LAE from the 
observations. \cite{carilli2013} tried to model the emission of the LAE taking into 
account the radiation coming from the luminous nearby quasar. All the models 
that reproduce the \cii and \lya luminosities predict higher luminosities for 
other UV lines that are not detected \citep{ohyama2004}.
Given this results, they conclude that the quasar is unlikely the source of heating 
and ionization in the LAE.

Based on deep, spatially resolved optical spectroscopy of the LAE, 
\cite{ohyama2004} argue that the LAE is likely the composition of a normal 
star-forming galaxy and an extended nebula with violent kinematic status. 
This nebula emission would produce a broadening of the \lya emission. This nebula 
can be explained, at least in a qualitative way, as a superwind caused 
by the supernovae explosion of  OB stars in the late phase of the evolution of a starburst.

In conclusion, this LAE is not intrinsically different from the general population of 
LAE. The \cii detection in this LAE can thus be used as a reference for 
\cii searches in other LAEs at high redshift.

\subsubsection{Himiko simulations.}
Simulations also suggest narrow \cii emission lines at high redshift for 
LAEs. \cite{vallini2013} combine a high resolution cosmological simulation with a 
sub-grid multi-phase model of the interstellar medium to simulate the \cii emission 
in a halo similar to the LAE Himiko at $z=6.6$. They find that 95\% of the 
\cii emission is generated in the Cold Neutral Medium (CNM), mainly in 
clumps of individual size $\leq 3$ kpc. They present a spectrum for the simulated 
\cii emission, where the FWHM of the main peak is $\sim 50$ \kms, very similar 
to the 56 \kms of the LAE at z=4.7. This suggests that the width of the \cii line is 
to first order determined by the gravitational potential of the clumps. The \cii emission 
produced in the CNM follows the gravitational potential of the clumps, resulting in narrow
emission lines associated with each clumps. An ensemble of emitting clumps 
moving through the galaxy following the potential of the galaxy could combine and produce a broader line. Such behavior 
is not observed in the simulations, where just a small number of clumps dominate the \cii emission.

We conclude that the adopted width of $\sim 50$ \kms for the \cii line in LAEs agrees
with recent observations and simulations. Nevertheless, we do not discard the possibility of \cii lines being
broader than our assumption, but that the occurrence of unusually narrow lines in this population appears
plausible.

\subsection{CMB effects.}
\label{sec:cmbeffects}
The Cosmic Microwave Background Radiation (CMB) emits as a black body with a temperature of
 \tcmbo$ =2.7$ K. The temperature of the CMB increases linearly with ($1+z$), becoming an 
important factor to take into account for observations of objects at high redshift. \cite{dacunha2013}  
showed the effect of the CMB on observations of high-redshift galaxies. Here we will follow the 
prescription formulated by \cite{dacunha2013} to take into account the effects of the CMB in the 
continuum observations of galaxies at high redshift in the mm and sub-mm. We will apply this 
prescription to the SED of the local galaxies, as if they would be observed at a given redshift z.

The templates that we use are those presented by  \cite{silva1998} for the galaxies Arp 220, M82, 
M51 and NGC 6946. For the galaxies assume cold dust with temperature \tdusto\ and an emissivity
index $\beta$. For Arp 220 we used $\tdusto =66.7$ K and $\beta=1.86$ \citep{rangwala2011}, for 
M82 $\tdusto =48$ K and $\beta=1$ \citep{colbert1999}, for M51 $\tdusto =24.9$ K 
and $\beta=2$ \citep{mentuchcooper2012} and for NGC 6946 we used $\tdusto =26$ K and 
$\beta=1.5$ \citep{skibba2011}. 
At a given redshift the CMB contributes to the dust heating such that the equilibrium temperature is:

\begin{equation}
\tdust(z) = \Big( (\tdusto)^{4+\beta} + (\tcmbo)^{4+\beta} \big[(1+z)^{4+\beta} - 1 \big] \Big) ^{\frac{1}{4+\beta}}\,.
\label{tdustz}
\end{equation}
$\tdusto$ is a measurement of the mean dust temperature as determined by a modified
blackbody fit to an observed galaxy IR SED at $z = 0$, representing the total IR luminosity of the galaxy. As a representative fit, this is equally applicable to both optically thin galaxies and optically thick as in Arp 220. So long as the galaxy is transparent to the CMB radiation (true for even Arp 220), Eq. 1 holds.
The additional heating by the CMB affects the SEDs such that the peak of the emission is shifted to a shorter wavelength and 
the total luminosity associated to the cold dust is higher by $[\tdust(z)/\tdusto]^{(4+\beta)}$. 
We need to modify the intrinsic SED of the galaxies to include this new $\tdust(z)$. The flux density 
depends on the black body radiation for the given temperature,

\begin{equation}
F_{\nu/(1+z)}\propto B_{\nu}(T_{dust}(z)),
\end{equation}

To include $\tdust(z)$ we have to apply the following factor to convert the intrinsic SED flux density to the emission associated with the new temperature $F^{*}_{\nu/(1+z)}$.

\begin{equation}
F^{*}_{\nu/(1+z)}=F^{\rm int}_{\nu/(1+z)}\times \frac{B_{\nu}(\tdust(z))}{B_{\nu}(\tdusto)}.
\end{equation}

This factor will only apply to the part of the SED that correspond to the emission of the cold dust. To accomplish this, we scale a modified black body (MBB) to the peak of the FIR emission of the SED at $\tdusto$, and then use this MBB emission to estimate the ratio ($R_\nu$) of emission associated with the cold dust at a given frequency,

\begin{equation}
R_{\nu}	=	\frac{K\nu^{\beta}B_{\nu}(\tdusto)}{F^{\rm int}_{\nu}},
\end{equation}
where $K$ is just the scaling factor. The flux density associated to the new temperature of the cold dust will be:

\begin{equation}
F^{*}_{\nu/(1+z)}=M_{\nu}\times F^{\rm int}_{\nu/(1+z)}
\end{equation}

with 

\begin{equation}
M_{\nu}=\left[\left(1-R_{\nu}	\right)+R_{\nu}	\times \frac{B_{\nu}(\tdust(z))}{B_{\nu}(\tdusto)}\right].
\end{equation}

Finally, following \cite{dacunha2013}, we have to take into account the effect of the CMB as an observing background. For this we have to multiply the flux associated
with $\tdust(z)$ by $C_{\nu}$,

\begin{equation}
C_{\nu}=\left[1-\frac{B_{\nu}(\tcmb(z))}{B_{\nu}(\tdust(z))}\right],
\end{equation}

resulting in the flux observed of the galaxies as:

\begin{equation}
F^{\rm obs}_{\nu/(1+z)}=C_{\nu}\times M_{\nu}\times F^{\rm int}_{\nu/(1+z)},
\end{equation}

with $C_{\nu}\times M_{\nu}$ representing the effect of the CMB in the observations at a given frequency. 
The same corrections are derived when an optically thick emission is assumed, as in the case of Arp 220.

As we can see in Fig. \ref{fig:SEDs}, the effect of the CMB decreases the observable flux density at 2 mm by up to a factor of $5\times$ (in the case of M51) for the galaxy at $z\sim 11$, when
the temperature of the CMB is higher, as expected. Also, the effect is higher for galaxies with a lower temperature of the cold dust.
Galaxies with temperature of order of 25-30 K are more affected than those with temperature of 40-50 K. 
The CMB effects will be important for estimations of the flux densities of these type of galaxies in the continuum and for 
the correct interpretation of the observations.

The CMB effects on the \cii line observations are similar to those on the continuum. 
The flux of an emission line observed against the CMB
is:

\begin{equation}
\label{eq:lineflux_aCMB}
\frac{S^{\rm obs}_{\nu/(1+z)}}{S^{\rm int}_{\nu/(1+z)}}=\left[1-\frac{B_{\nu}(\tcmb(z))}{B_{\nu}(T_{\rm exc})}\right],
\end{equation}

with $T_{\rm exc}$ being the excitation temperature of the transition.
For the case of local thermal equilibrium (LTE), when collisions dominate the excitation of the \cii line, the excitation temperature of the transition 
is equal to the kinetic temperature of the gas ($T_{\rm kin}$). The kinetic temperature varies for the
different \cii emission regions. Gas temperatures within PDRs are typically  $T\sim100-500$ K \citep{stacey2010},
for the CNM  $T\approx 250$ K, for the WNM $T\approx 5000$ K and for the ionized medium $T\approx 8000$ K\citep{vallini2013}.
Since the CMB temperature at $z=6.5-11$ is much lower than the gas temperature of the \cii emitting region, it will not contribute significantly to the \cii excitation, but must be taken into account as the background against which the line flux is measured. In most of the \cii emission regions, the temperatures
are so high that the observed flux of the line against the CMB is similar to the intrinsic flux ($S^{\rm obs}_{\nu/(1+z)}/S^{\rm int}_{\nu/(1+z)}\approx 1$). For the extreme case where all the \cii emission is being produced in PDRs with temperature of 100 K in a galaxy at $z=11$, the 
observed flux (using Eq. \ref{eq:lineflux_aCMB}) would be 90\% of the intrinsic flux. 
We found this case very unlikely, since in low redshift galaxies the \cii emission produced in PDRs is 50-70\% of the total \cii luminosity and the gas temperatures associated to the PDRs are higher \citep{crawford1985,carral1994,lord1996,colbert1999}. 
We conclude that the CMB effects on the \cii line observations are negligible for our observations.

\subsection{Spectral energy distribution of the galaxies.}
\label{sec:sed}
Using the upper limits on the continuum, we compare the targets with 
the spectral energy distribution templates of local galaxies (SEDs). For the LAEs, the SEDs 
of the local galaxies are scaled to the flux of a near-IR filter that is not contaminated 
by the \lya emission line. For \lbg, the filter used for the scaling is the one next 
to the Lyman Break. The photometry of  IOK-1 and \lbg together with the SED 
of  local galaxies are shown in Fig. \ref{fig:SEDs}. 
For \sdf1 and \sdfa (not shown) the situation is very similar: the sources have a similar redshift, 
the continuum upper limits are comparable and the CMB effects are of the same order.
Our upper limit for IOK-1 is 
comparable to the upper limit found by \cite{walter2012} using PdBI observations. 

Using the SED of NGC 6946 as a template, we estimate 
the IR luminosity given the upper limit flux densities, similar to the approach shown by 
\cite{walter2012}. We scale the SED of NGC 6946 to the 3 sigma upper limits of 
the mm observations and integrate from 8 $\mu\rm m$ to 1 mm (rest frame) to 
compute the IR luminosity.

The IR luminosity corresponding to this intrinsic SED and the SFR 
associated \citep{kennicutt1998} are given in Tab. \ref{tab:table1} for the LAEs and 
in Table \ref{tab:table2} for \lbg.
We note that estimating the IR luminosity using NGC 6946
without taking into account the CMB result in a significant underestimation  of the luminosity upper limits. 
The IR luminosity limit corrected by the CMB  of 
the LAEs at $z\sim 6.6$ is 35\% higher than without correcting by the CMB. For 
IOK-1 at $z\sim7$, the IR luminosity limit is a 50\% higher than the estimation without 
correcting by the CMB. For \lbg at $z\sim10.7$, the IR luminosity limit corrected 
for the CMB is $\sim 3.5$ times the IR luminosity limit not corrected by the CMB. 
For galaxies with cold dust temperature of $\sim 25$ K, the effect of the 
CMB on the observations is very important at high redshift, and it will significantly limit the 
feasibility of detecting not extremely starbursting galaxies in the IR continuum, it will not greatly affect the detectability of \cii emission.

\subsection{Ratio $L_{\cii}/L_{\rm FIR}$}
\begin{figure}
\epsscale{1.2}
\plotone{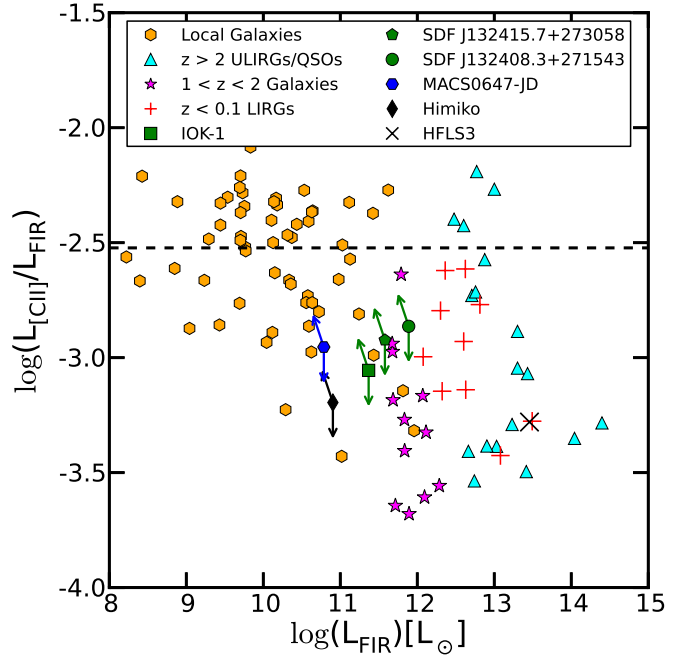}
\caption{Ratio of the \cii luminosity to the FIR luminosity vs the FIR luminosity (integrated from $42.5 \mu{\rm m}$ to $122.5 \mu{\rm m}$ rest frame) for galaxies at different redshifts.
The green symbols correspond to the upper limits of the LAEs presented here. The blue hexagon corresponds to the upper limit of \lbg using the 
most sensitive setup. The FIR luminosities for the galaxies are upper limits estimated from the observations including the CMB effects. The black diamond corresponds to the upper limit for Himiko with ALMA observations \citep{ouchi2013}. 
The horizontal dashed line is the average value for $L_{\cii}/L_{\rm FIR}$ on the local galaxies.
\citep{malhotra2001,negishi2001,luhman2003,iono2006,maiolino2009,walter2009,stacey2010,
ivison2010,wagg2010,cox2011,debreuck2011,swinbank2012,venemans2012,walter2012a,wang2013,riechers2013,ouchi2013}
\label{fig:CIIFIR}}
\end{figure}

Figure \ref{fig:CIIFIR} presents our upper limits to $L_{\cii}/L_{\rm FIR}$ and $L_{\rm FIR}$ together
with detections of \cii in other galaxies. The arrows represent the region of possible values for 
$L_{\cii}/L_{\rm FIR}$ and $L_{\rm FIR}$ (integrated from $42.5 \mu{\rm m}$ to $122.5 \mu{\rm m}$ rest frame). If we used UV-based SFR estimates to infer $L_{\rm FIR}$, 
our data points would move across the diagonal arrows towards the region where local galaxies are, 
putting our $L_{\cii}/L_{\rm FIR}$ upper limits close to the average value found
for the local galaxies. 
The ratio $L_{\cii}/L_{\rm FIR}$ is a measure of how efficient the \cii emission is
in cooling the gas. 
The values presented for our targets, $\log(L_{\cii}/L_{\rm FIR})\sim-2.9$, do not necessarily imply 
that \cii is not efficient in cooling the gas in these galaxies, it is most likely a consequence of the galaxies having 
much lower FIR luminosities than our conservative upper limits. 
Different processes can affect the ratio $L_{\cii}/L_{\rm FIR}$.
In galaxies with low extinction and low metallicity, like in Haro 11, about 50\% of the \cii emission arises from the diffuse ionized 
medium \citep{cormier2012}. Variations on the fraction of \cii emission associated with the ionized medium will also affect the ratio $L_{\cii}/L_{\rm FIR}$.
In some galaxies, the internal dust extinction can affect the ratio $L_{\cii}/L_{\rm FIR}$. In Arp 220, the dust is optically thick at 158 $\mu$m and 
can absorb part of the \cii emission, decreasing the ratio $L_{\cii}/L_{\rm FIR}$ \citep{rangwala2011}.

\cite{diaz-santos2013} present the results of a survey of \cii in luminous infrared galaxies (LIRGs)
observed with the PACS instrument on board the \textit{Herschel Space Observatory}. They found a tight correlation
between the ratio $L_{\cii}/L_{\rm FIR}$ and the far-IR $S_{\nu}(63\mu m)/S_{\nu}(158\mu m)$ continuum color, independently
of their $L_{\rm IR}$. They found that the ratio decreases as the average temperature of dust increases, suggesting that the main observable linked to the variation of  $L_{\cii}/L_{\rm FIR}$ is the average dust temperature.
For the galaxies with dust temperatures $\sim20 K$ the average ratio is $\log(L_{\cii}/L_{\rm FIR})\sim 10^{-2}$, suggesting that for galaxies like NGC 6946 with a dust temperature of $\approx26$ K, the ratio $L_{\cii}/L_{\rm FIR}$ should be on the same order of magnitude.
\cite{diaz-santos2013} also found a correlation between $L_{\cii}/L_{\rm FIR}$ and luminosity surface
density of the mid-IR emitting region ($\Sigma_{\rm IR}=L_{\rm IR}/\pi r^{2}_{\rm mid-IR}$). LIRGs with
lower $L_{\cii}/L_{\rm FIR}$ ratios are warmer and more compact. We can use this relation to find a rough estimation for $L_{\cii}/L_{\rm FIR}$ of our targets.
As $r_{\rm mid-IR}$ we use the size found in the UV observations of the targets. The half-light radius of IOK-1 is $\approx0.62$ kpc \citep{cai2011}.
The full width at half maximum size of \sdf1 and \sdfa  are $\approx 4.0$ and 3.2 kpc respectively \citep{taniguchi2005}. For \lbg  the delensed
half-light radius is $\lesssim 0.1$ kpc \citep{coe2013}. Using our $L_{\rm IR}$ upper limits as an approach to $L_{\rm IR}$ we can estimate $\Sigma_{\rm IR}$. For IOK-1 the estimated ratio is $\log(L_{\cii}/L_{\rm FIR})\sim-2.9$,  for \sdf1 is $\log(L_{\cii}/L_{\rm FIR})\sim-2.5$ and for \sdfa is $\log(L_{\cii}/L_{\rm FIR})\sim-2.6$. For the LAEs the average of $L_{\cii}/L_{\rm FIR}$ is similar to the average value for the local galaxies (Fig. \ref{fig:CIIFIR}).
For \lbg the estimated ratio is $\log(L_{\cii}/L_{\rm FIR})\sim-3.2$.

\subsection{SFR-$L_{\cii}$ Relation}
\begin{figure}
\epsscale{1.2}
\plotone{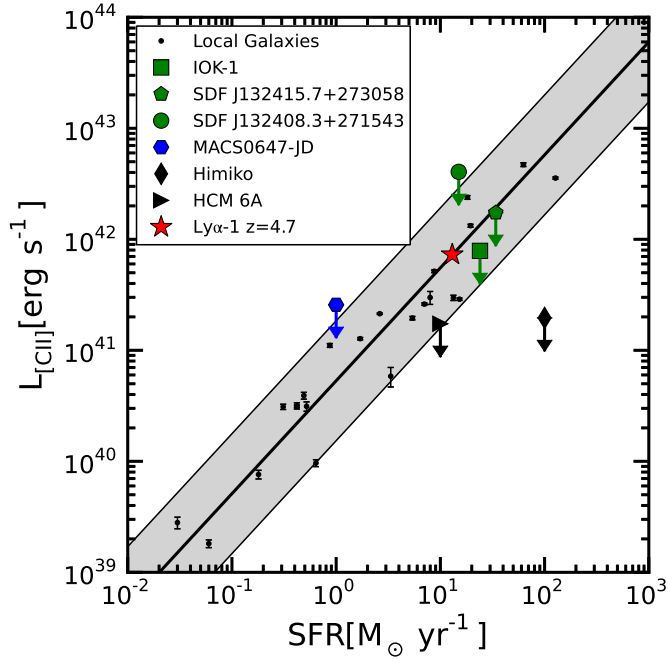}
\caption{Relation of the \cii luminosity with the UV-derived star formation rate of galaxies. The black solid lines correspond to the relation
found by \cite{delooze2011}, with the gray area corresponding to $2\sigma$ of the scatter in the relation. The black dots with error bars
correspond to the data used to find the relation of \cii - SFR. The green circle, square and pentagon correspond to the LAEs with the \cii upper limits presented in this paper
assuming the star formation rate estimated from the UV fluxes. The blue hexagon corresponds to the \cii upper limit of \lbg with based in the most sensitive setup and the star formation rate estimated from the UV fluxes. The red star corresponds to the LAE detected with ALMA at $z\sim4.7$ \citep{carilli2013}. The black triangle corresponds to the upper limit of the \cii emission found for HCM-6A by \cite{kanekar2013}.  The black diamond corresponds to the upper limit of the \cii emission found for Himiko by \cite{ouchi2013}. \label{fig:CIISFR}}
\end{figure}

In Fig. \ref{fig:CIISFR} we present our $L_{\cii}$ upper limits with the UV-SFR estimated for the targets together with upper
limits detections for published LAEs \citep{carilli2013,kanekar2013,ouchi2013}. The black solid lines corresponds to the relation
found by \cite{delooze2011}, with the gray area corresponding to $2\sigma$ scatter in the relation. Our upper limits 
for the \cii luminosity fall within the scatter of the SFR-$L_{\cii}$, with the exception of \sdfa, where the upper limit falls above the
relation due to the moderate depth of its observations. The detection of the LAE at z=4.7 (\lya-1) agree very well with the relation found by \cite{delooze2011} using the UV-SFR estimated by
\cite{ohyama2004}. The upper limits for the lensed LAE at $z=6.56$ HCM 6A and Himiko suggest that LAEs at $z>6$ 
could fall below the relation found at low redshift. More observations are needed to clarify if there is an intrinsic difference between
the LAEs at $z\sim4.5$ with the higher redshift population. 
The high magnification of \lbg allows us to explore an UV-SFR one order of magnitude lower than the ones of the LAEs, showing the advantage
of observing lensed galaxies to cover the intrinsically faint population at high redshift.

\subsection{IOK-1 Models}
\label{sec:iok1_simulations}
\begin{figure}
\epsscale{1.2}
\plotone{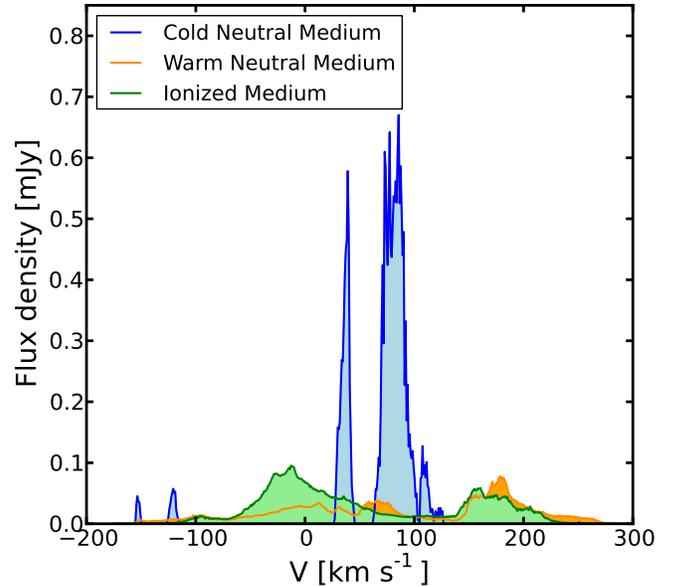}
\caption{Simulated \cii spectrum of a galaxy similar to IOK-1 at $z\sim7$. The parameters set for this simulation
 were a SFR of $20$M$_{\odot}$yr$^{-1}$, and stellar population age of 10 Myr and a solar metallicity. The blue spectrum corresponds
 to the emission produced in the cold neutral medium, the orange spectrum corresponds to the emission produced in the warm neutral medium and 
 the green spectrum corresponds to the emission produced in the ionized medium. The main peak (at $\sim 80 \kms$)  of the cold neutral medium has a FWHM of $\sim50$ \kms. For more details on the simulations of \cii emission in high redshift galaxies see \cite{vallini2013}.
 \label{fig:specCIIZ1}}
\end{figure}

Using the same procedure presented in \cite{vallini2013} for the \cii emission of Himiko, we estimate the emission of \cii for IOK-1 at $z\sim7$.
 For this simulation, the star formation rate was set to 20 $\msol yr^{-1}$ and a stellar population age of 10 Myr. The metallicity was set to solar to have
 a conservative estimation of the \cii emission. The simulation does not include the emission from PDRs and should be seen as a lower limit. In Fig. \ref{fig:specCIIZ1}, we show the \cii emission produced by the three modeled phases, cold neutral medium (CNM), warm neutral medium (WNM) and the ionized medium. Most of the \cii emission comes from the CNM ($\sim50\%$), the rest
 is coming from the WNM ($\sim20\%$) and from the ionized medium ($\sim30\%$). For comparison, in Himiko,  95\% of the emission is produced in the CNM and the rest in the WNM. No emission from the ionized medium was modeled in the simulation of Himiko \citep{vallini2013}. We can also see in the emission that the FWHM of the main peak is $\sim 50$ km\ts s$^{-1}$, just as expected. 
 
\begin{figure}
\epsscale{1.2}
\plotone{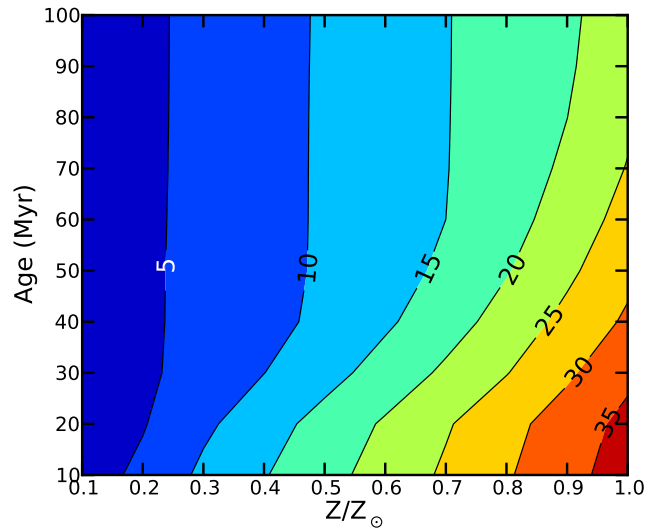}
\caption{Contour plot of the integrated \cii flux of IOK-1 in ${\rm mJy} \kms$ for different simulation conditions. As comparison, our upper limit for integrated flux of IOK-1 is 175 ${\rm mJy} \kms$. The two independent parameters are the stellar population 
age and the metallicity of the gas. The flux is integrated over the whole area of the cube and in a channel
resolution of 500 \kms around the peak of the emission. The integrated flux is a conservative upper limit for the different parameters.
We can see from the contour plot that the \cii emission is very sensitive to the metallicity of the galaxy, and in a less significant way to the age of the stellar population. 
The different ages correspond to a different amount of heating photons coming from the young stars, which is critical for the cooling of the gas.
\label{fig:contour_iok1}}
\end{figure}

In Fig.  \ref{fig:contour_iok1}, we present the integrated flux of \cii for a different combination of metallicities and stellar population ages. This shows a strong dependency on the metallicity, which is expected, since it is treated linearly with the abundance of \cii in the gas.
The second main feature of this results is the dependency with the stellar population age. Here we assumed a continuum star formation rate of 20 $\msol yr^{-1}$, for the older stellar populations there is a higher amount of heating photons coming from the UV part of the spectrum. This is a result of using a continuum star formation mode, for a given SFR, older populations have more time generating young UV emitting stars. These extra heating photons avoid the cooling of the gas, which decrease the amount of gas in the cold neutral medium, where most of the \cii emission is produced.

\subsection{Spectral Resolution}

For a Gaussian emission line, with a FWHM of 50 \kms observed at a channel resolution of 200 km\ts s$^{-1}$, emission lines will be significantly diluted.  In the best case scenario of the 
line falling completely in one channel, we will recover 38\% of the peak flux density of the line. This suggests to carry out observations a sufficiently high 
spectral resolution. E.g. with a line of FWHM of 50 \kms and a channel resolution of 10 km\ts s$^{-1}$, we expect to recover 
97\% of the peak flux density of the line.

\subsection{Atomic Mass Estimation}
We use Equation 1 from \citep{hailey-dunsheath2010} to give rough upper limits to the atomic mass associated with PDRs in our targets (Assuming all \cii would arise from PDRs). As approach to the PDRs conditions we use the result of \cite{vallini2013} for the temperature and density in the CNM of Himiko,
$n=5\times10^{4} $ cm$^{-3}$ and T=250 K. Using our upper limits for \cii we obtain the following upper limits to the atomic mass:
For IOK-1 $M_{\hi}\lesssim 2\times10^{8}\msol$, for \sdf1 $M_{\hi}\lesssim 4\times10^{8}\msol$, for \sdfa $M_{\hi}\lesssim 1\times10^{9}\msol$ and
for \lbg $M_{\hi}\lesssim 6\times10^{7}\msol$.
Assuming that the mass of atomic gas is similar to the mass of molecular gas, we can compare our upper limits with the measurements of similar galaxies at lower redshift.

The only molecular gas masses measured in high redshift UV-selected star-forming galaxies come from the detection of CO transition lines in lensed LBGs. 
The measured values are, $\sim 4\times10^{8}\msol$, $\sim 9\times10^{8}\msol$ and $\sim 1\times10^{9}\msol$ for  MS 1512--cB58 ($z=2.73$), the cosmic eye ($z=3.07$)
and MS1358--arc ($z=4.9$) respectively \citep{coppin2007, riechers2010, livermore2012}. Our upper limits for the LAEs are similar to the values estimated for the observed LBGs. For \lbg our upper 
limit for the molecular mass is at least $8\times$ lower than in the observed LBGs.

Using the UV-SFR relation we can estimate the gas depletion timescales for our targets, assuming $\tau_{\rm dep}=M_{\rm gas}/{\rm SFR_{\rm UV}}$. We estimate upper limits for the depletion time of
$\lesssim 8 \rm Myr$, $\lesssim 11 \rm Myr$ and $\lesssim 66 \rm Myr$ for IOK-1, \sdf1 and \sdfa respectively. For \lbg the depletion time is $\lesssim 60 \rm Myr$. The estimated depletion times
for the observed lower redshift lensed LBGs are within the range of $\sim7-24$ Myr, similar to our upper limits. The depletion times for the LAEs are consistent with the ages estimated for the young population of LAEs 
at $z\sim4.5$ of $<15$ Myr found by \cite{finkelstein2009} and to the simulated LAEs at $\sim3.1$ with ages $<100$ Myr \citep{shimizu2011}. The depletion times of the LBGs are consistent with the
LBG-phase predicted duration of $20-60$ Myr \citep{gonzalez2012}.

\cite{saintonge2013} presented molecular gas masses and depletion timescales for a sample of lensed star forming galaxies at $z=1.4-3.1$. The range of measured molecular gas masses is $5.6\times10^{9}-4\times10^{11}\msol$ and of depletion timescales is $127-1089$ Myr. 
The longer depletion timescales measured for the lower-z sources could indicate that they experience less `extreme' bursts of star formation in comparison to our $z>6.5$ sample. Although, assuming a higher molecular-to-atomic gas ratio (of at least 5) would put our upper limits within the values measured by \cite{saintonge2013}.

\section{Summary and Outlook.}

We have presented a search for \cii emission in three LAEs 
at $z\sim 7$ and in a LBG at $z\sim11$ using CARMA and the PdBI. 
We summarize our results and conclusions as follows:

\begin{enumerate}
\item We have not detected \cii emission line any of our targets. 
Given the recent observational results and simulations of the \cii emission in high redshift LAE, we adopt
 a line width of 50 \kms for the \cii emission. We put constrains on the luminosity of the line for the targets.
 For the LAEs the $3\sigma$ L$_{\cii}$ upper limit are $<2.05$, $<4.52$ and $<10.56\times10^{8}\lsun$ for
 IOK-1, \sdf1 and \sdfa respectively. Our \cii upper limits are consistent with the relation of SFR-L$_{\cii}$ found by \cite{delooze2011}.
 The $3\sigma$ upper limit in the \cii luminosity of \lbg is  $<5.27\times10^{7}\times(\mu/15)^{-1}\lsun$ (Assuming that the redshift of the galaxy is within the most sensitive setup).

\item No detection of the FIR continuum is found at a wavelength of 158 $\mu \rm m$ rest frame for any of the 4 targets.
Assuming a spectral energy distribution template for the local galaxy NGC 6946 as a template for the high redshift galaxies observed here, 
we present conservatives upper limits for the FIR luminosity.  We find $<2.33$, $3.79$ and $7.72\times10^{11}\lsun$ as upper limits
for IOK-1, \sdf1 and \sdfa respectively, these values account for the effect of the CMB on the observations.
For MACS0647-JD, the upper limit in the FIR luminosity is $<6.1\times10^{10}\times(\mu/15)^{-1}\lsun$, after correcting for the CMB and the lensing magnification.

\item We present the results of simulations supporting the brightest component of the \cii line having a width of the order of $50$ \kms. Here we want to emphasize the necessity of resolving such emission lines in future ALMA observations, to not loose signal-to-noise ratio, by selecting a channel resolution that is too low. 
 
\item The effect of the CMB must to be taken into account in attempts to detect the FIR continuum in galaxies at high redshift. 
The heating of cold dust by CMB photons can shift the peak of the FIR continuum to values up to a $\sim400$ microns for 
galaxies with temperature of $\sim25$ K and redshift of $z\sim11$. 
We emphasize that not including the effects of the CMB on the observations results in an underestimation of the FIR luminosities for the targets.
The CMB corrected FIR luminosity limits are 35\% higher than those without CMB correction at $z\sim6.6$, 50\% higher at $z\sim7$, and 350\% higher
at $z\sim11$ for a T$=26$ K.

\item Simulations are already showing us that the task of detecting \cii in high redshift galaxies is going to be difficult even with ALMA, as confirmed
by the recent sensitive non-detection of Himiko by \cite{ouchi2013}. 
Accordingly to our IOK-1 simulations, a key parameter for the \cii emission in LAEs is the metallicity, as we discussed in Sect. \ref{sec:iok1_simulations}. 
If these simulations were applicable to all high redshift LAEs, we should first try to detect \cii in the LAEs with the highest metallicity.
Estimating the metallicity of LAEs at high redshift is not an easy task, however, \cite{cowie2011} found that for the sample of LAEs discovered by the Galaxy 
Evolution Explorer (GALEX) grism in the redshift range of $z=0.195-0.44$, there is an anti-correlation of the equivalent width of the H$\alpha$ emission line with metallicity. Higher EW(H$\alpha$) sources all have lower metallicities, bluer colors, smaller sizes, and less extinction. \cite{cowie2011} also found a broad general trend that for higher EW(H$\alpha$), the EW($\rm{ly}\alpha$) is also higher. If we assume that these relations are valid for the LAEs at high redshift, and that the goal is to observe the LAE with the highest metallicity possible, it may be best to target the brightest LAE in the UV but with the lowest \lya equivalent width. Lyman-break galaxies with \lya detection may thus be ideal targets for \cii searches at high redshift. 
\end{enumerate}

We thank the referee for his/her useful comments and suggestions which significantly improved the quality of this paper. We thank Brent Groves for the discussion on the effects of the CMB.
Support for CARMA construction was derived from the states of California, Illinois, and Maryland, the James S. McDonnell Foundation, the Gordon and Betty Moore Foundation, the Kenneth T. and Eileen L. Norris Foundation, the University of Chicago, the Associates of the California Institute of Technology, and the National Science Foundation. Ongoing CARMA development and operations are supported by the National Science Foundation under a cooperative agreement, and by the CARMA partner universities.
Based on observations with the IRAM Plateau de Bure Interferometer. IRAM is supported by INSU/CNRS (France), MPG (Germany) and IGN (Spain).
LI and JG obtained partial support from CATA, Conicyt Basal program. LI and JG acknowledge support from FONDAP ``Centro de Astrof\'isica" 15010003. LI thanks the collaboration of the CLASH team.

\clearpage

\end{document}